\documentclass[conference]{IEEEtran}
\IEEEoverridecommandlockouts

\usepackage{amsmath,amssymb,amsfonts}
\usepackage{hyperref}
\usepackage{romannum}
\usepackage{algorithmic}
\usepackage{graphicx}
\usepackage{textcomp}
\usepackage{xcolor}
\usepackage{comment}
\usepackage{booktabs}
\def\BibTeX{{\rm B\kern-.05em{\sc i\kern-.025em b}\kern-.08em
    T\kern-.1667em\lower.7ex\hbox{E}\kern-.125emX}}
  
\usepackage{multirow}
\usepackage[caption=false]{subfig} 
\usepackage{enumitem}

\usepackage{flushend}

\newcommand{\state}[1]{\mathcal{S}_{#1}}
\newcommand{\PWM}[0]{\mathtt{PWM}}
\newcommand{\sel}[0]{\mathtt{sel}}
\newcommand{\trigger}[2]{\mathcal{T}_{#1 \rightarrow #2}}

\graphicspath{{./figures/}}

\begin{document}

\title{A Control Philosophy for Multiplexed Power Converters in Active Distribution Systems\\
\thanks{M. Deakin was supported by the Royal Academy of Engineering under the Research Fellowship programme. This work was supported by the Engineering and Physical Sciences Research Council (EPSRC) Impact Acceleration Account (IAA) at Newcastle University, EP/X525601/1. Contact: \texttt{matthew.deakin@newcastle.ac.uk}}
}

\author{\IEEEauthorblockN{Matthew Deakin}
\IEEEauthorblockA{\textit{School of Engineering} \\
\textit{Newcastle University}\\
Newcastle-upon-Tyne, UK}
\and
\IEEEauthorblockN{Mahmood Jamali}
\IEEEauthorblockA{\textit{School of Engineering} \\
\textit{Newcastle University}\\
Newcastle-upon-Tyne, UK}
\and
\IEEEauthorblockN{Nail Tosun}
\IEEEauthorblockA{\textit{School of Engineering} \\
\textit{Newcastle University}\\
Newcastle-upon-Tyne, UK}
\and
\IEEEauthorblockN{Shafiq Odhano}
\IEEEauthorblockA{\textit{School of Engineering} \\
\textit{Newcastle University}\\
Newcastle-upon-Tyne, UK}
}

\maketitle

\begin{abstract}
Reconfigurable, multiplexed power electronic devices (M-PEDs) can provide more effective, flexible operation for applications in active distribution networks, enabling the mitigation of thermal constraints and voltage violations. To-date, operational approaches of these M-PEDs have not been described, a significant issue as a result of their relative complexity as compared to conventional PEDs. This work proposes a state machine-based control philosophy that can be used to manage the mutually exclusive selector- and current-control signals for M-PEDs. Two control approaches are described using this philosophy: an Off-Load approach, suitable for systems without time-critical network constraints; and a Hot-Swap approach, implemented when the impact of transients must be minimized. Simulations illustrate the evolution of the state machine for a back-to-back voltage source converter topology. It is concluded that the proposed state machine-based control philosophy enables M-PED operation whilst managing the additional complexity of this promising converter topology.
\end{abstract}


\section{Introduction}

Power electronic devices (PEDs), such as back-to-back soft open points (SOPs) and four-wire static compensators (4W-STATCOMs), are becoming a prevalent solution for distribution system operators (DSOs) to address voltage violations and thermal constraints, aiming to increase capacity headroom or reduce curtailment of distributed energy resources (e.g., rooftop solar). To increase PED flexibility, reconfigurable multiplexed PEDs (M-PEDs) have been proposed as a PED topology that offers increased flexibility when the PED operates as a multiport converter. In particular, M-PEDs include a switching matrix for reconfiguration (i.e., a \emph{power multiplexer}, p-mux) at the output of each of the power converter's modules. The inclusion of these p-muxes can increase the flexibility of the PED by a factor of two or greater (in terms of peak power transfer per port), depending on the converter topology and application \cite{deakin2023multiplexing,cui2023two,deakin2024reconfigurable}.

\begin{figure}\centering
\includegraphics[width=0.49\textwidth]{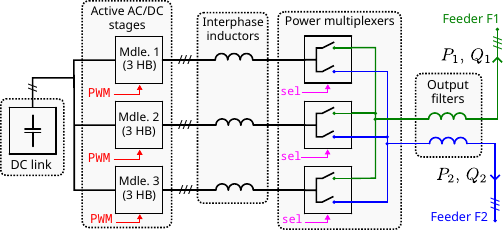}
\caption{This work proposes a control philosophy that defines how current control signals, $\PWM$, and multiplexer selector signals, $\sel$, are coordinated to enable a transition between configurations and power injections into each feeder (i.e., the feeder to which each module (Mdle.) is connected). The example M-PED we consider in the remainder of the paper is a 3-Module, 2-Feeder M-SOP.}\label{f:reconfig-topology-control}
\end{figure}

One PED type which can benefit from this structure is a back-to-back soft open point (SOP). An example multiplexed SOP (M-SOP) topology is shown in Fig.~\ref{f:reconfig-topology-control}. Compared to conventional (non-reconfigurable) SOPs, the operating condition of an M-PED is defined not only by pulse-width modulation signals ($\PWM$, defining the active and reactive powers $P_k,\,Q_k$ injected by the $k$th module), but also by the state of the multiplexer's selector signals ($\sel$, defining which output port that power is transferred to or from).

For conventional (non-reconfigurable) PEDs, proportional-integral (PI) control of the outer voltage loops and inner current loops is dominant for grid-following PEDs such as SOPs \cite{jiang2022overview} or static compensators (D-STATCOMs) \cite{hashemzadeh2021secondary}. In contrast, there are very few time-domain operational studies considering a PED's changing quasi-static converter topology, with a notable study considering a PED's bypass switch in \cite{lai2023transient}. In contrast, there are increasing numbers of papers considering M-PED operation that focus on quasi-steady-state (QSS) network-level optimal operation \cite{deakin2023multiplexing,cui2023two}, where the output of such techniques are the active and reactive injections of each module together with multiplexer configuration. Similar QSS studies have been considered for other related configuration problems, such as (modular) fast charging station distribution systems \cite{saner2024social}, reconfigurable ac/dc substations \cite{shekhar2020offline}, or network configuration for fixed PED operation in hybrid ac/dc systems \cite{ahmed2019energy}.

Given the significant increase in complexity of M-PEDs compared to conventional PEDs, the lack of time-domain control approaches suitable for describing the M-PED operation is a substantial gap. In particular, this gap leaves uncertainty as to how M-PED operational complexity might be addressed reliably and efficiently, as required to realise the benefits of M-PEDs for DSOs. The contribution of this work is to address this gap by proposing a complete control philosophy for M-PEDs, covering QSS operation (through a network-level control system) through to the generation of $\PWM$ signals and multiplexer switch $\sel$ signals. Simulations results for an M-PED prototype validate the effectiveness of the proposed control philosophy.

This work is structured as follows. We first outline the DSO's requirements for both the PED and M-PED control in an electrical distribution network (Section~\ref{s:control_requirements}). The proposed state machine-based M-PED control philosophy is then introduced, with two control approaches compared against classical PI control to demonstrate the philosophy's effectiveness (Section~\ref{s:transients}). The implementation of the proposed approach is considered through a hypothetical time-domain evolution of the state machine, as it could be implemented for prototype hardware (Section~\ref{s:experiments}). Finally, conclusions are drawn (Section~\ref{s:conclusion}).

\section{PED and M-PED Control Requirements for Operation in Active Distribution Networks}\label{s:control_requirements}

PEDs (and M-PEDs) are installed by DSOs to increase the capability of the network to deliver power to end-customers safely and reliably. The purpose of their installation could typically be expected to increase network capacity by managing thermal constraints or voltage violations; furthermore, they must do so without compromising the quality of power supplied to the end customer---i.e., without compromising electromagnetic compatibility (EMC). If a PED is used to mitigate voltage violations, then it implies that the PED must have the capability to significantly impact on network voltages at its PCC; therefore, EMC with regard to transient- and steady-state voltages must be considered.

The most relevant EMC constraints for PED and M-PED operation relate to voltage sags and swells and rapid voltage change (RVC). For a concrete consideration of these issues, we consider EMC requirements for the Great Britain (GB) region. Although the specific thresholds used to define these issues vary \cite{barros2021review}, this is considered a reasonable region to be representative for considering (M-)PED operational requirements.

Voltage sags and swells are measured by power quality monitors when rms voltages rise above or below a threshold. This means a PED or M-PED should ideally avoid (through its action or inaction) causing the voltage $V_t$ to go outside the acceptable limits
\begin{equation}\label{e:swell}
V^{\mathrm{Sag}} \leq V_t \leq V^{\mathrm{Swell}}\; \forall \, t\,,
\end{equation}
where $V_t$ is the full-cycle rms voltage measured at time $t$, updated every half-cycle (10~ms); and $V^{\mathrm{Sag}},\,V^{\mathrm{Swell}}$ are the sag and swell voltage thresholds, respectively. The severity of a sag/swell can be classified depending on changes in the duration of the voltage excursion \cite{british2025voltage}; this is not considered further in this work.

RVC constraints (or, equivalently, \emph{step voltage constraints}) mean that changes in voltage magnitude should not exceed an upper threshold, $\mathrm{RVC}^{\mathrm{Max.}}$, i.e.,
\begin{equation}\label{e:rvc_definition}
\left | V_\mathrm{PCC}(S^{\mathrm{PED,\,new}}) - V_\mathrm{PCC}(S^{\mathrm{PED,\,old}}) \right | \leq \mathrm{RVC}^{\mathrm{Max.}}\,,
\end{equation}
where $S^{\mathrm{PED,\,new}},\,S^{\mathrm{PED,\,old}}$ are hereon referred to as \emph{Old} and \emph{New} (M-)PED setpoints, and $V_\mathrm{PCC}$ is the voltage at the point of common coupling (PCC). The value of $\mathrm{RVC}^{\mathrm{Max.}}$ is typically 3\%~pu \cite{barros2021review}. If voltage changes occur over a short time period or are infrequent, then more complex RVC envelopes can be defined that are less conservative; nevertheless, \eqref{e:rvc_definition} is considered the most appropriate operational constraint for M-PEDs and PEDs considered in this work.

If the PCC voltage changes by more than 3\% as a result of the change in setpoints, then a PED or M-PED can be operated so that its output is sufficiently slow to not introduce an RVC constraint \cite{ena2026erep_p28}. The \emph{steady state voltage} is defined within the appropriate standard \cite{ena2019p28} as a 1~s period over which the voltage does not change by more than 0.5\%. Whilst a maximum voltage slew rate for $V_{\mathrm{PCC}}$ is not presented directly in the standards \cite{ena2026erep_p28,ena2019p28}, this steady state voltage definition does imply that inequality~\eqref{e:rvc_definition} can be neglected as a constraint if instead
\begin{equation}\label{e:rvc_implied}
\dfrac{\mathrm{d} V_\mathrm{PCC}}{\mathrm{d} t} \leq 0.5 \% /\mathrm{s}\,.
\end{equation}

\subsection{Conventional PED Control to Meet EMC Requirements}\label{ss:ped_control}

The control of conventional PEDs is typically maintained through a hierarchical control structure. A network-level control system (NLCS) determines active and reactive power setpoints, then passes these to a real-time PED controller that ultimately generates PWM signals to realise the reference power injections. These PWM signals are typically generated from the output of PI controllers which track reference d-q current signals within a cascaded voltage control loop \cite{jiang2022overview,hashemzadeh2021secondary}. 

\begin{figure}
\centering
\includegraphics[width=0.22\textwidth]{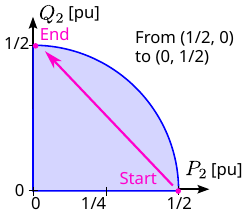}
\caption{The first quadrant of the capability chart for a SOP in active and reactive powers, $P_2,\,Q_2$, respectively. SOP capability charts are convex, so all points between two feasible setpoints are also feasible.}\label{f:philosophy-conventional}
\end{figure}

For a conventional PED, the voltage sag/swell constraint \eqref{e:swell} can be implicitly accounted for by lower-level PI controllers. This is because the operating region is convex (see Fig.~\ref{f:philosophy-conventional}) and, assuming a linear relationship between PCC voltages and d-q currents (as considered in appropriate examples of \cite{ena2026erep_p28}), then all operating points between the New and Old NLCS setpoint will be feasible, not passing through any regions which would result in a voltage sag or swell. If RVC is a relevant concern, then the slew rate constraints \eqref{e:rvc_implied} can be implemented by choosing a time constant which is sufficiently large for the relevant control loops, such that inequality~\eqref{e:rvc_definition} can be neglected.

\subsection{Characterising M-PED Control Philosophy Challenges}\label{ss:mped_challenge}

\begin{figure}
\subfloat[Example configuration (i)]{\includegraphics[width=0.46\textwidth]{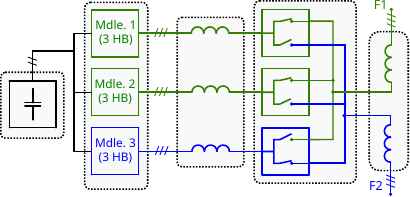}\label{ff:config1}}\\
\subfloat[Example configuration (ii)]{\includegraphics[width=0.46\textwidth]{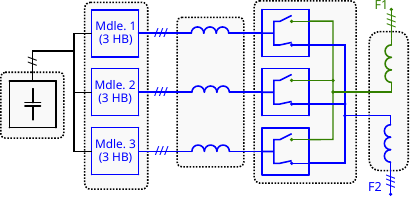}\label{ff:config2}}
\caption{Two example M-SOP configurations, (i), (ii) that can be realised by changing the M-SOP's selector signals $\sel$.}\label{f:reconfig-topology-configurations}
\end{figure}

In general, an M-PED will need to change not only the PWM signals $\PWM$ for each converter, but also the configuration of the power multiplexers through $\sel$ (Fig.~\ref{f:reconfig-topology-control}). For example, two configuration states of that M-SOP are shown in Fig.~\ref{f:reconfig-topology-configurations}.

In contrast to a PED, there is considerable non-convexity in an M-PED's operating region \cite{deakin2022design}, in terms of the powers that can be injected into each feeder. This non-convexity means that New and Old setpoints will not, in general, have a continuous operating trajectory between them. This means that a given transition path of the M-PED's current injections may result in a sag or swell. Furthermore, there is a need to change the discrete values of power multiplexer configuration signals $\sel$.

There are two approaches that could address this issue. The first would be to aim to complete a transition from the old setpoint to a new setpoint within 10 ms (i.e., half a cycle). This would mean that the rms measurement of voltages by a power quality monitor would not register the issue. This has several substantial practical challenges:
\begin{itemize}
\item it will rule out electromechanical switches as a means of implementing the multiplexers (as proposed as an efficient candidate solution for implementation \cite{deakin2022design}), due to their settling times of tens of ms (e.g., 25~ms operating time for an 80~A contactor \cite{sensata2026minitactor});
\item either precise zero-current switching will be required to be implemented for the multiplexer switches, or substantial snubbers applied to avoid large voltage transients during on-load switching of the multiplexer's series switches;
\item potentially large angle differences between the feeders (or phases) which are connected for the Old and New PED setpoints create a risk of `shoot-through', so an appropriate deadband should be applied between opening a switch connected to one feeder and closing another;
\item if PI controllers are used for current control, these will need very high gains so that new steady-state sinusoidal currents can rapidly be achieved;
\item If the voltage changes more than 3\%, then a one-shot change from an Old to New setpoint within 10~ms would violate \eqref{e:rvc_definition}, so a series of setpoint changes would need to be defined (noting, if this is the case, then if the series of setpoints is not chosen appropriately, there could be intermediate points within this trajectory for which voltage sag/swell conditions may not be met).
\end{itemize}
Considering these challenges, it is considered the complexity of such an approach will be impractical for a typical M-PED practical, including the experimental implementation developed within this work which uses electromechanical contactors (Section~\ref{s:experiments}).

Therefore, for the remainder of this work, we focus on a second approach. This assumes a relatively slow transition through M-PED configurations. This approach can be has constraints on the controller design.
\begin{enumerate}
\item[(i)] the individual switches within the power multiplexer can only be switched off-load. This enables the control to be used with a multiplexer constructed from either electromechanical- or solid-state switches.
\item[(ii)] The values of $\sel$ should never permit more than one multiplexer switch to be closed at any one time.
\end{enumerate}
A corollary of (i), (ii) and the non-convexity of M-PED's operating regions is that it may be impossible for the M-PED controller to avoid all transient sags and swells when the M-PED is being used to avoid voltage violations. We therefore aim to provide a framework which can be the basis of \emph{minimizing} the impact of these setpoint transitions aiming to avoid (wherever possible) voltage sags/swells \eqref{e:swell} whilst respecting the RVC ramp-rate constraint \eqref{e:rvc_implied}. Finally, we add the following requirement,
\begin{itemize}
\item[(iii)] the operation of an M-PED should aim to be simple and understandable, where this is possible.
\end{itemize}
This final requirement (iii) should be considered heuristic, and can be dropped if the control designer cannot clearly differentiate between different algorithms.

\section{An M-PED Control Philosophy}\label{s:transients}

The power multiplexer switches can only be switched when they are in an off-load condition (i). This means that each M-PED module's $\PWM$ and $\sel$ control signals are mutually exclusive: either the output current can be controlled, or the state of the multiplexers can be changed. These mutually exclusive control signal pairs lend themselves naturally to a state machine-based control approach. Appropriate logic checks can be implemented within the state machine to ensure that no more than one switch within a multiplexer is closed at any one time (ii).

\begin{figure}
\includegraphics[width=0.49\textwidth]{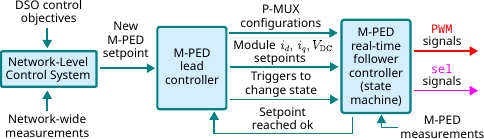}
\caption{Proposed hierarchical M-PED control architecture as a means of implementing the controller requirements.}\label{f:overall-philosophy}
\end{figure}

The proposed control philosophy has a three-stage architecture, as shown in Fig.~\ref{f:overall-philosophy}. The lowest-level M-PED follower controller implements the state machine to enable or disable $\sel$ or $\PWM$ signals, with conventional PI controllers implemented for individual module current and voltage control loops, tuned to satisfy \eqref{e:rvc_implied}. This is driven by a lead M-PED controller whose function is to convert a New M-PED setpoint, defined as the active and reactive powers injected into each feeder, into a series of d-q currents, power multiplexer states, and state transition trigger signals that will result in transitioning the M-PED from an initial setpoint to a new setpoint whilst meeting \eqref{e:swell} where possible. The M-PED's per-feeder power injection setpoint has been determined by the network-level control system using a DSO's operational objectives and network-wide measurements. The implementation of the network-level control system has been previously studied, for example, as constrained optimization problems, and is not elaborated on further (see, e.g., \cite{deakin2023multiplexing,deakin2022design}).

\subsection{M-PED Follower Control State Machine}

The M-PED follower controller uses a conventional control approach to realise stable operation, utilising per-module P-Q and Vdc-Q modes during normal operation, driving each module's $\PWM$~signal \cite{cao2016operating}. Additional states are then defined when some or all output $\PWM$ signals are suspended, as necessary to enable the $\sel$~signal for one or more modules. (Other states that are also commonly implemented, such as PED start-up sequences, are not considered in this work.)

Trigger signals are implemented to indicate when the follower controller should transition to a new state. Guard conditions are implemented, checking when the trigger is received that the module is in an appropriate condition prior to transitioning to a new state. Firstly, a module transitioning from $\PWM$ to $\sel$ control must have zero current on all phases (to within a numerical tolerance), to meet (i). Conversely, if a module will transition from $\sel$ to $\PWM$ control, it is checked that the controller has a fixed multiplexer configuration (e.g., sufficient time required to operate any contactors has passed).

Under this general state machine-based control philosophy, two specific control approaches are considered for the follower controller.
\begin{itemize}
\item \emph{Off-Load} control, with two states. Either all modules have $\PWM$ enabled and are conducting, with no capabilities to change configuration ($\state{0}$); or, all modules have $\sel$ enabled and are nonconducting, with no possibility to inject current ($\state{1}$). A pair of triggers $\trigger{0}{1},\,\trigger{1}{0}$ indicate when the controller should transition between $\state{0}$ and $\state{1}$.
\item \emph{Hot-Swap} control, with $n+1$ states for an $n$-module M-PED. The first state, $\state{0}$, has all modules conducting with $\PWM$ enabled; thereafter, the $(k+1)$th state $\state{k+1}$ corresponds to the $k$th module nonconducting (with $\sel$ enabled) and remaining modules conducting with $\PWM$ enabled. Triggers $\trigger{0}{k},\,\trigger{k}{0}$ enabling a transition to and from state $\state{k}$ from $\state{0}$ respectively, and there are therefore $2n$ triggers in total.
\end{itemize}
For both of the control approaches, one module is chosen to operate under a Vdc-Q control mode, with all other modules chosen with a P-Q setpoint, both implemented using appropriate PI controllers as in \cite{cao2016operating}. An example of these two control approaches is summarised in Table~\ref{tb:states-off-load} and Table~\ref{tb:states-hot-swap} for the M-SOP of Fig.~\ref{f:reconfig-topology-control}. 

\begin{table}
\centering
\caption{Off-Load control capabilities for each state.}
\label{tb:states-off-load}
\begin{tabular}{rccc}
    \toprule
    State & Module 1 & Module 2 & Module 3 \\
    \midrule
    $\state{0} $ & $\PWM$ (P-Q) & $\PWM$ (P-Q) & $\PWM$ (Vdc-Q) \\
    $\state{1} $ & $\sel$ & $\sel$ & $\sel$ \\
    \bottomrule
\end{tabular}
\end{table}

\begin{table}
\centering
\caption{Hot Swap control capabilities for each state.}
\label{tb:states-hot-swap}
\begin{tabular}{rccc}
    \toprule
    State & Module 1 & Module 2 & Module 3 \\
    \midrule
    $\state{0} $ & $\PWM$ (P-Q) & $\PWM$ (P-Q) & $\PWM$ (Vdc-Q) \\
    $\state{1} $ & $\sel$ & $\PWM$ (P-Q) & $\PWM$ (Vdc-Q) \\
    $\state{2} $ & $\PWM$ (P-Q) & $\sel$ & $\PWM$ (Vdc-Q)\\
 	$\state{3} $ & $\PWM$ (P-Q) & $\PWM$ (Vdc-Q) & $\sel$\\
    \bottomrule
\end{tabular}
\end{table}

\subsubsection{Implementation of an M-PED lead controller}

Depending on the state machine implementation, the complexity of the M-PED lead controller varies. For an Off-Load control approach, the lead controller can be simple. Current setpoints are first set to zero, then the trigger $\trigger{0}{1}$ moves the state to $\state{1}$, disabling $\PWM$ and enabling $\sel$. Each multiplexer is set to its new configuration, then trigger $\trigger{1}{0}$ re-enables $\PWM$ and disables $\sel$, after which the d-q current setpoints are chosen to realise the M-PED setpoint. If there are no voltage sag / swell issues of concern, then this control is considered appropriate to meet the simplicity requirement (iii).

There is more complexity in the M-PED lead controller implementation for the Hot Swap approach, although the approach may be necessary to meet voltage sag/swell and RVC requirements \eqref{e:swell}, \eqref{e:rvc_implied}. In general, the overall transition between M-PED setpoints can be implemented as a series of transitions through states---the current should be set to zero for one module, with the current injections of the other modules chosen to achieve appropriate operational objectives (e.g., ensuring \eqref{e:swell}, \eqref{e:rvc_implied} are met). Once the multiplexer configuration has changed, the configuration of each other module is then also updated in-turn in the same manner, until the M-PED reaches the New setpoint.

The operating point chosen during Hot Swap control will significantly influence whether a network meets its operational objectives. Furthermore, the preceding states change the capability chart when a module reaches state $\state{k}$ (due to a different multiplexer configuration), and therefore the order through which states are passed can also change the characteristics of the operational change. Future work could consider optimal approaches to choose these transitions so that the risk or severity of sags or swells is mitigated.

\subsection{Example of Off-Load and Hot Swap Control Approaches}\label{ss:control_examples}

\begin{figure}\centering
\subfloat[Off-Load approach]{\includegraphics[height=0.28\textwidth]{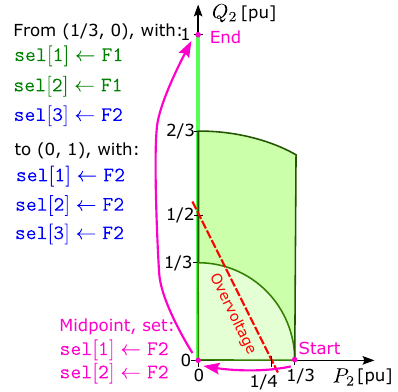}\label{ff:philosophy-offload}}~~~
\subfloat[Hot-Swap approach]{\includegraphics[height=0.28\textwidth]{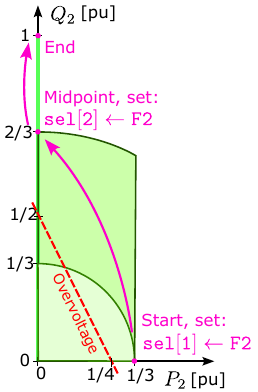}\label{ff:philosophy-hotswap}}~
\caption{Comparison of the two control approaches for the M-SOP (only showing the first quadrant). Compared to the SOP (Fig.~\ref{f:philosophy-conventional}), the M-PED has increased control complexity, but with increased flexibility in terms of reactive power injections.}\label{f:philosophy}
\end{figure}

To illustrate both M-PED control approaches and contrast these with conventional PED operation, we consider an M-PED with M-SOP topology shown in Fig.~\ref{f:reconfig-topology-control}, comparing this against a conventional SOP with 1/2 pu module capacity connected to two feeders. For the M-SOP, we consider an Old setpoint change from an active power injection of 1/3 pu from feeder F1 to F2 to New setpoint as a fully reactive injection of 1 pu only on feeder F2 (Fig.~\ref{f:philosophy}). The configuration of the multiplexers should change so that the 1/3 pu module capacity initially connected to feeder F2 is increased to 1 pu, as shown in Fig.~\ref{f:reconfig-topology-configurations}. For the SOP, we consider a change from an active power injection of 1/2 pu from F1 to F2 to a fully reactive injection of 1/2 pu on F2, as shown in Fig.~\ref{f:philosophy-conventional}. 

Comparing the conventional SOP control (Fig.~\ref{f:philosophy-conventional}) and proposed M-SOP control approaches (Fig.~\ref{f:philosophy}), the complexity of changing the M-PED setpoint is clear. Nevertheless, the additional power injection that is possible with the M-PED is also evident, with a doubling of the reactive power that can be injected (on a per-unit basis).

Comparing Fig.~\ref{ff:philosophy-offload} and Fig.~\ref{ff:philosophy-hotswap}, the additional complexity of the Hot Swap control can be seen as it must change state four times (changing multiplexer configurations first at the Start, and then at the Midpoint), as compared to just twice for the Off-Load control. Nevertheless, the requirement to pass through the zero-load point is also a distinct disadvantage of Off-Load control, resulting in a transient overvoltage, thereby failing to meet the sag and swell constraint \eqref{e:swell}.

\begin{figure}\centering
\subfloat[Off-Load M-PED control]{
	\includegraphics[width=0.45\textwidth]{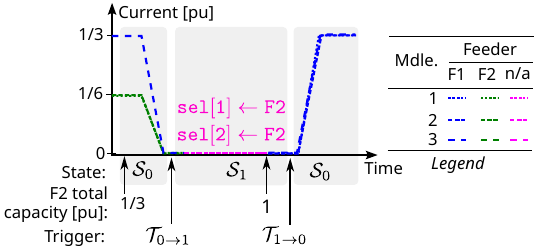}
	\label{ff:time-offload}
}\\
\subfloat[Hot-Swap M-PED control]{
	\includegraphics[width=0.45\textwidth]{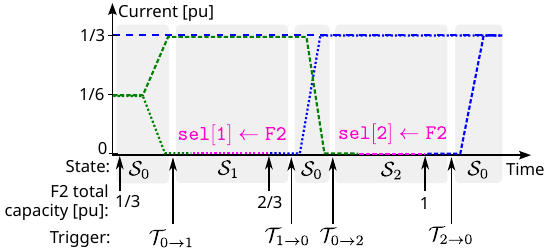}
	\label{ff:time-hotswap}
}
\caption{Time-domain summary of states and transitions for the Hot-Swap and Off-Load control approaches for the changes in configuration and injections, considering an M-SOP topology (Fig.~\ref{f:reconfig-topology-control}).}\label{f:time-domain}
\end{figure}

The timing of trigger and multiplexer configuration signals for the M-SOP are shown in Fig.~\ref{f:time-domain}. An implementation of Off-Load control can be developed using a simple subroutine to implement the steps shown in Fig.~\ref{ff:philosophy-offload}. For the Hot Swap control, the controller must choose active and reactive setpoints that satisfy guard conditions and simultaneously meet sag and swell constraints \eqref{e:swell}, as well as determining the order of state transitions.

\section{Simulation Results}\label{s:experiments}

\begin{figure}\centering
\includegraphics[width=0.35\textwidth]{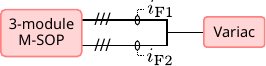}
\caption{The grid is simulated as two feeders, fed from a variac to represent a three-phase ac source, connected to the three-module reconfigurable M-SOP. The currents in feeders F1 and F2 are denoted $i_\mathrm{F1},\,i_\mathrm{F2}$ respectively.}\label{f:test-setup}
\end{figure}

To consider the practicality of the proposed M-PED control philosophy, a simulation of a three-module M-SOP has been developed (Fig.~\ref{f:reconfig-topology-control}) to demonstrate time-domain evolution of the proposed state machine. The power multiplexer's electromechanical relays, which must be switched off-load, are modeled as ideal switches with a small parallel impedance which acts as a snubber (which is necessary to avoid numerical issues during simulation). The switches are assumed to have a current rating of 10~$\mathrm{A_{rms}}$, with the per-unit current base 30~$\mathrm{A_{rms}}$. The grid is modeled as a three-phase 50~V rms voltage simulating an ac distribution grid with two feeders, F1 and F2, as shown in Fig.~\ref{f:test-setup}. Currents are contolled with conventional dq axis current control using PI controllers. Simulations are conducted in PLECS.

\begin{figure}\centering
\includegraphics[width=0.5\textwidth]{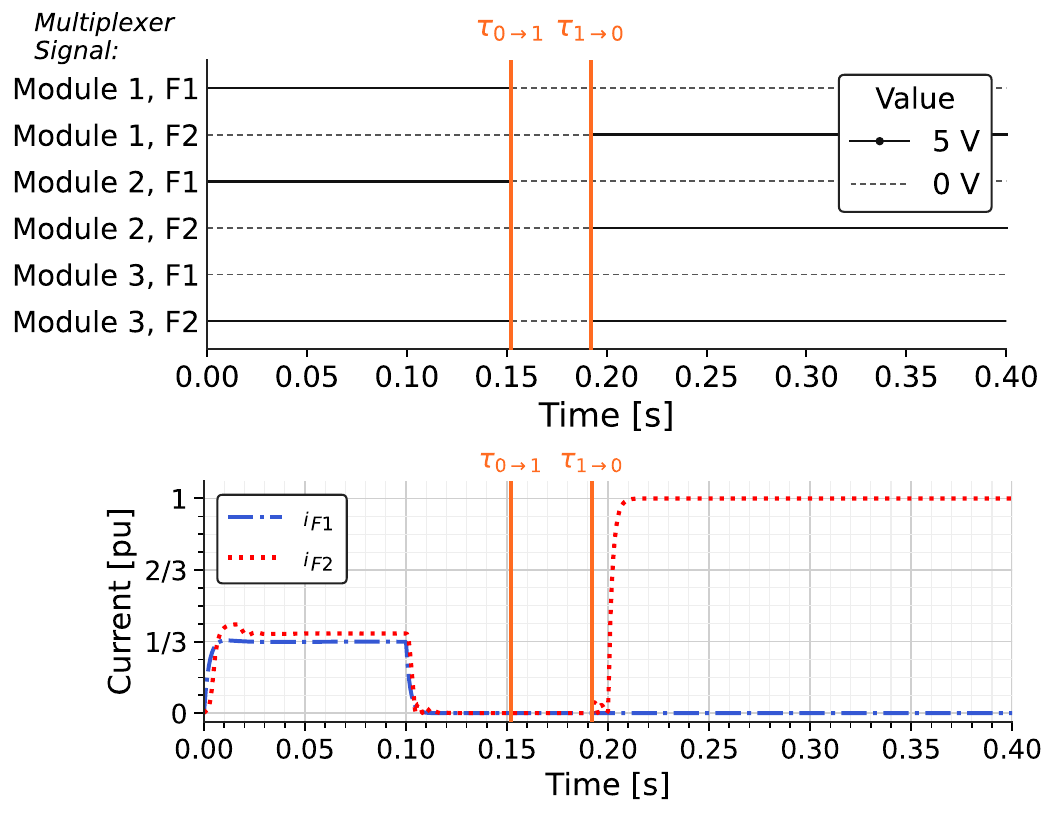}
\caption{Simulation results for the Off-Load control approach. The subfigures show the multiplexer coil energization (upper), representing state of the power multiplexer; and the rms feeder current (lower). Both subfigures show the timing of the trigger signals to request the M-PED follower control to change state.}\label{f:example-results-off-load}
\end{figure}

For the Off Load approach, the transition signals follow the approach described in Section~\ref{ss:control_examples}. Fig.~\ref{f:example-results-off-load} shows the state of the multiplexer matrix (via the energization of the contactor coils) and the current injected into each feeder. It can be seen that the changes in the energization occur only between trigger signals $\trigger{0}{1},\,\trigger{1}{0}$ (i.e., when the M-SOP is in state $\state{1}$), and that current is only non-zero outside of this period (i.e., in state $\state{0}$).

\begin{figure}\centering
\includegraphics[width=0.5\textwidth]{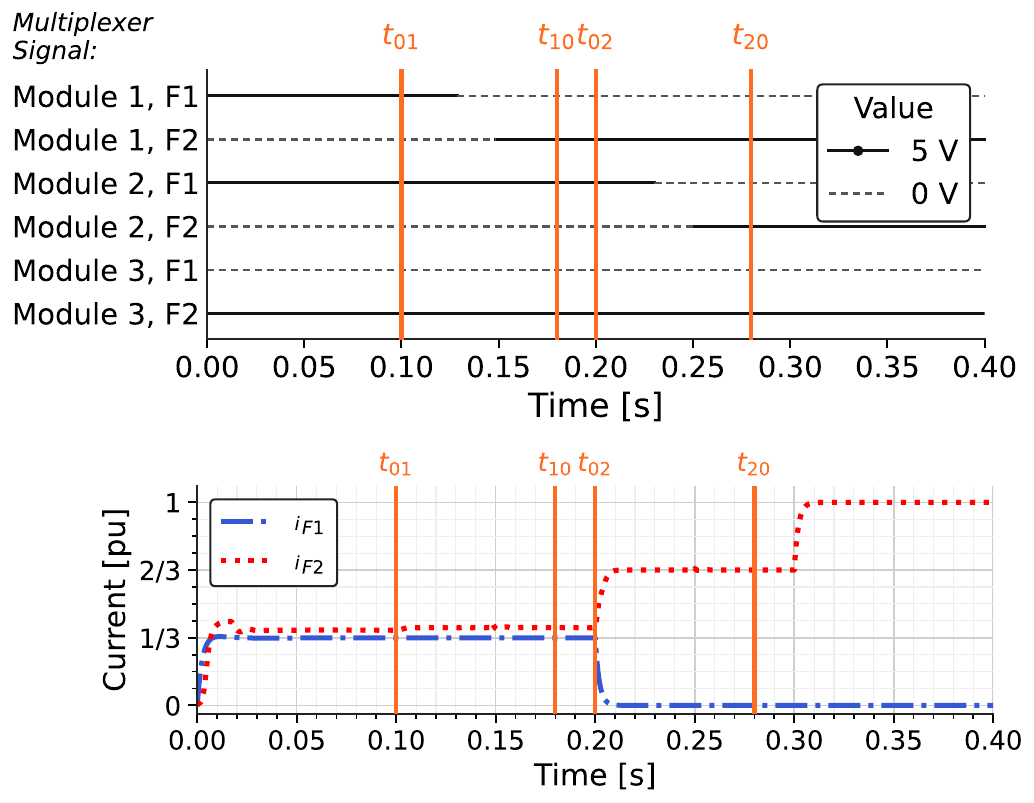}
\caption{Simulation results for the Hot Swap control approach, demonstrating the ability to change the configuration of an M-PED whilst continuing to inject some power into the distribution network.}\label{f:example-results-hot-swap}
\end{figure}

The time-domain state machine evolution for the Hot Swap approach also follow the signals introduced in Section~\ref{ss:control_examples}. Fig.~\ref{f:example-results-hot-swap} shows the state of the multiplexer matrix and the injections of feeder current alongside the trigger signals. It can be observed that the reconfiguration of the $k$th module only occurs between the trigger signal pair $\trigger{0}{k},\,\trigger{k}{0}$ (i.e., in state $\state{k}$). As with the Off-Load approach, a change in configuration of the multiplexer for a given module occurs when that module is not conducting (the implemented current setpoints follow the magnitude of the per-module injection, as shown in Fig.~\ref{ff:time-hotswap}). However, in contrast to the Off-Load approach, the Hot Swap approach continues to inject current during the transition to the new PED setpoint.

Note that the initial difference between the feeder RMS currents is due and have a current rating of 10~$\mathrm{A_{rms}}$. A three-phase 50~V rms voltage can be to the DC-link voltage control action, since the converter regulating the DC link injects the additional active power required to balance the DC-link energy. This behaviour appears in both control examples and is not caused by the reconfiguration process itself.

The requirements for off-load switching (i) and allowing only one selector output at a time (ii) have been met for both control approaches. In the example shown, the variac provides a stiff ac grid at the PCC of the M-SOP, which means that the requirements \eqref{e:swell} and \eqref{e:rvc_definition} are trivially met. Future experimental works could include feeder impedances to demonstrate that the Hot Swap approach can avoid transient voltage sags and swells. If an M-SOP were implemented in a real distribution grid with a stiff voltage source, its simplicity implies that the Off Load approach would be preferred (by (iii)).

\section{Conclusions}\label{s:conclusion}

To enable the benefits of reconfigurable M-PEDs to be realised, control approaches must be developed that can address their increased operational complexity as compared to conventional PEDs. This work has developed a control philosophy to address this issue, proposing to take existing network-level control M-PED setpoints as the control input and then using this to determine PWM and multiplexer configuration signals through the implementation of a state machine. The state machine ensures that mutually exclusive reconfiguration and current control signals cannot occur simultaneously. The scalability of the architecture is clear with the number of states growing only linearly with M-PED modules in the worst case.

The Off-Load approach provides a clear and simple implementation of the proposed control philosophy, although it risks transient voltage sag or swell events due to the no-load condition during multiplexer reconfiguration. In contrast, the Hot Swap approach sequentially reconfigures M-PED's multiplexers, enabling the mitigation of voltage changes as the M-PED changes its setpoint, albeit with an increase in complexity. Whilst the Hot Swap approach does not afford the continuously feasible operation of a conventional PED, it does offer the potential to minimize the severity and duration of voltage sags and swells. Future work could further explore M-PED lead controller implementation to develop approaches which achieve this objective through systematic optimization approaches.

Reconfigurable power converters have been proposed for diverse applications, from information technology equipment power supplies to electric vehicle fast charging stations. By accounting for the power distribution network requirements in the development of the proposed control, we conclude that M-PEDs have the potential to provide an effective and efficient solution to meet multiport-type mission profiles commonly seen in distribution power flow control applications.


\end{document}